# Design and demonstration of an acoustic right-angle bend

Wenjia Lu[1], Han Jia[1,2], Yafeng Bi[1], Yuzhen Yang[1] and Jun Yang[1,2]

[1] Key Laboratory of Noise and Vibration Research, Institute of Acoustics, Chinese Academy of Sciences, Beijing 100190, People's Republic of China

[2] State Key Laboratory of Acoustics, Institute of Acoustics, Chinese Academy of Sciences, Beijing 100190, People's Republic of China

E-mail: hjia@mail.ioa.ac.cn;   jyang@mail.ioa.ac.cn

## Abstract

In this paper, we design, fabricate and experimentally characterize a broadband acoustic right-angle bend device in air. Perforated panels with various hole-sizes are used to construct the bend structure. Both the simulated and the experimental results verify that acoustic beam can be rotated effectively through the acoustic bend in a wide frequency range. This model may have potential applications in some areas such as sound absorption and acoustic detection in pipeline.

# 1. Introduction

Recently, the technique of transformation optics was introduced to establish a correspondence between material constitutive parameters and coordinate transformations [1, 2]. It is based on the invariance of Maxwell's equations under coordinate transformations [3]. The direct current conductivity equation was also shown to be invariant as a corollary of the invariance of Maxwell's equations [4]. Furthermore, researchers found that there is a one-to-one correspondence between the acoustic wave equation and the direct current conductivity equation [5]. Due to the invariance of the acoustic wave equation under coordinate transformations [6], the general principle of transformation acoustics is established and applied to both two dimensions (2D) and three dimensions [5-8].

Transformation acoustics provides an unprecedented ability to manipulate and control the behavior of sound wave, which includes cloaking, omnidirectional absorber, acoustic gradient-index lens, extraordinary transmission and subwavelength imaging [5, 6, 9-12]. The device design is obtained via a spatial coordinate transformation from the original free space to the transformed space, where the compression and dilation of space in different coordinate directions are interpreted as appropriate scaling of the material parameters (i.e. the mass density tensor and the bulk modulus). However, in most cases, these parameters are very complex: the space-dependent mass density and bulk modulus are usually inhomogeneous and extremely anisotropic. Although these parameters are challenging to achieve in practice, it becomes possible to realize the anisotropic desired parameters with the development of the metamaterial [13-17].

As an application of transformation theory, the right-angle bend has been proposed and experimentally demonstrated in optics [18, 19]. While expanding to acoustics, it can also be obtained using transformation acoustics. The simulated results are reported in our previous work [20]. In this paper, we design, fabricate and experimentally characterize an acoustic right-angle bend device in air. The quasi-two-dimensional device, which is made up of perforated plates with various hole-sizes [21-26], has surprising low complexity. The measurements of the acoustic field distribution around the bend are carried out inside a parallel plate waveguide. Its good performance demonstrates that the acoustic right-angle wave bend can work stably in a wide frequency range.

## 2. Theories and simulations

Our model is a 2D acoustic bend structure with an angle of $\pi/2$. The coordinate transformation is illustrated in figure 1. A square of size $w \times w$ in virtual space $(x, y)$ is shown in figure 1(a), which is transformed into a fan-shaped section in physical space $(r', \varphi')$ as shown in figure 1(b).

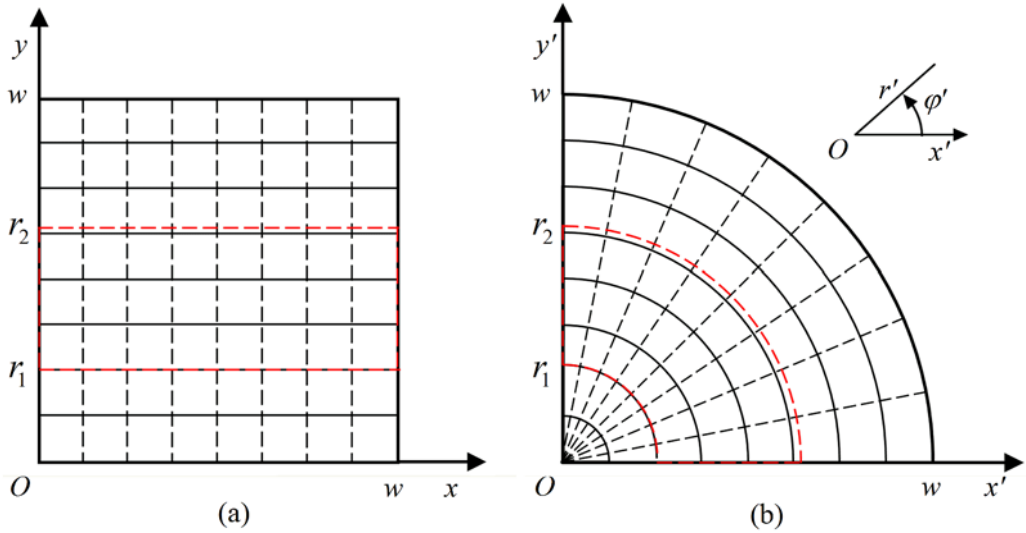

Figure 1. Spatial coordinate transformation used for the 2D right-angle bend model: (a) the original virtual coordinate space $(x, y)$; (b) the transformed physical coordinate space $(r', \varphi')$. The red dashed frames indicate the operation area for the proposed prototype.

The coordinate transformation between virtual space and physical space can be written as

$$r' = y, \quad \varphi' = \frac{\pi}{2w}(w - x) \tag{1}$$

where $r' = \sqrt{x'^2 + y'^2}$ and $\varphi' = tan^{-1}(y'/x')$. The Cartesian coordinates in the transformed systems can be retrieved from $x' = r'cos\varphi'$ and $y' = r'sin\varphi'$. From equation (1) and figure 1, it is obvious that constant $x$ lines are mapped into constant $\varphi'$ radial lines, while constant $y$ lines are mapped into constant $r'$ curves. If an incoming plane wave impinges normally on the interface $x=0$, in virtual space, the contours of equal phase are the parallel lines that are vertical to the $x$-axis, shown as the dashed lines in figure 1(a). After the transformation, the contours of equal phase change into the radial dashed lines shown in figure 1(b). The contours of equal phase are always perpendicular to the propagating direction. As a result, the propagating direction of the acoustic beam will be gradually rotated in physical space, finally to 90 degrees.

According to transformation acoustics [5-8], the spatial coordinate transformation from the original free space to the transformed space can be interpreted as appropriate scalings of the material parameters. The material parameters, i.e., the mass density tensor $\rho'$ and the bulk modulus $K'$ are given by

$$\frac{1}{\rho'(x',y')} = \frac{JJ^T}{\det J} \frac{1}{\rho_0(x,y)}$$

$$K'(x',y') = \det J\, K_0(x,y) I \tag{2}$$

where $\rho_0 = 1.21 kg/m^3$ and $K_0 = 0.14 MPa$ are the mass density and the bulk modulus of air. $I$ is the identity matrix. The elements of the Jacobian matrix $J$ are expressed as

$$J_{i,j} = \frac{\partial x_i'}{\partial x_j}, \quad i,j = 1,2 \tag{3}$$

By using equations (1-3), the material parameters of the bend can be obtained:

$$\rho_r' = \frac{\pi r'}{2w}\rho_0, \quad \rho_\varphi' = \frac{2w}{\pi r'}\rho_0$$

$$K' = \frac{\pi r'}{2w} K_0 I \tag{4}$$

where the parameters in the radial and angular directions are marked with subscript $r'$ and $\varphi'$ respectively.

Due to $c = \sqrt{K\rho^{-1}}$, we can additionally obtain the distribution of the sound speed expressed as $c_r' = c_0, c_\varphi' = \pi r' c_0/2w$, where $c_0$ is the sound speed of air. It can be observed that the sound velocity $c_\varphi'$ increases with the increase of the radius. The fact that the outer velocity is much faster than the inner velocity in this model leads to a capability of bending a propagating sound beam through a right angle bend. However, when $r'$ decreases to zero, $c_\varphi'$ will approach to zero. And when $r'$ is larger than $2w/\pi$, $c_\varphi'$ will be also larger than $c_0$. The materials with these acoustic velocities are difficult to realize in air. To avoid the extreme material parameters, we just choose a ringlike part of this model, which is indicated by the dashed red frame in figure 1(b). The inner radius $r_1$ is $4w/5\pi$ and the outer radius $r_2$ is $2w/\pi$. The corresponding area in virtual space is also indicated by the dashed red frame in figure 1(a).

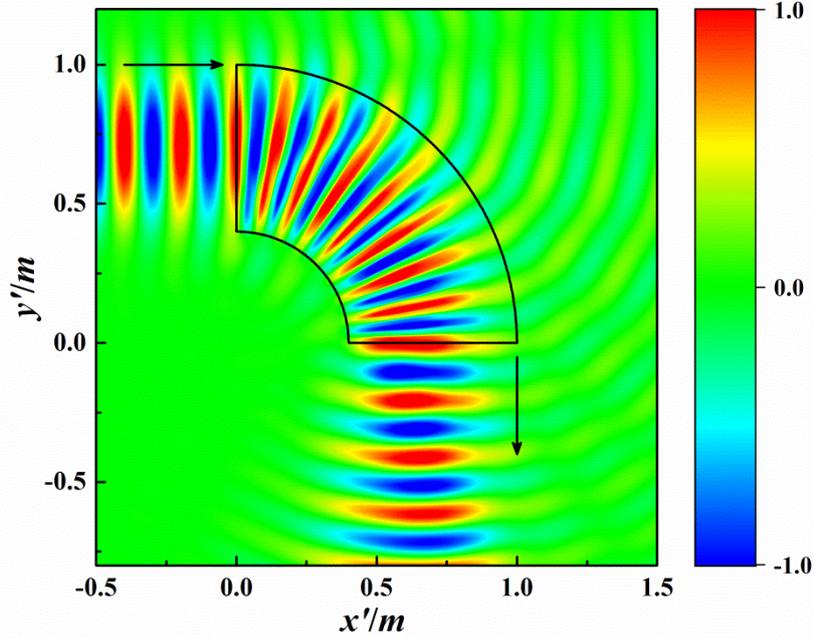

Figure 2. The normalized sound pressure field with a Gaussian beam incident on the right-angle bend at 1700 Hz.

Then a prototype is designed to validate the bending effect. Here we choose $w = \pi/2$ m as the length of the square in figure 1(a). Therefore, in our ringlike model mentioned above, the inner radius is $r_1 = 0.4$ m and the outer radius is $r_2 = 1$ m. Plugging $w = \pi/2$ m into equation (4), the required acoustic parameters of the bend can be obtained. Then the performance of the proposed model is simulated in frequency domain using finite element method (FEM). In simulation, the anisotropic acoustic parameters are achieved by layers of isotropic materials which linearly expand in the radial direction based on Biot fluid theory [16]. The boundaries of the sound field are set as absorbing boundaries to eliminate reflected waves. A Gaussian beam propagating in the $+x'$-direction is employed as the incident field to demonstrate the performance of the right-angle bend. As shown in figure 2, a Gaussian beam with frequency 1700 Hz is incident on the right-angle bend. The black arrows represent the propagating direction. It is clearly observed that the propagation direction of the incident Gaussian beam is gradually rotated by 90 degrees to the $-y'$-direction after passing through this right-angle bend. Here, the sound pressure field is normalized by the maximum value.

## 3. Design and demonstration

The simulated results have shown the possibility to realize the acoustic right-angle bend. Then, we will use acoustic metamaterial structure to realize the right-angle bend. Owing to $\boldsymbol{n} = c_0\sqrt{\boldsymbol{\rho K^{-1}}}, z = \sqrt{\boldsymbol{\rho K}}$, equation (4) above can be written as

$$n_r' = 1, \ n_\varphi' = r'^{-1}$$

$$z_r' = r'z_0, \ z_\varphi' = z_0 \qquad (5)$$

where $\boldsymbol{n}$ is the refractive index and $z$ is the acoustic impedance. Besides, $z_0 = 412\,N\cdot s/m^3$ is the acoustic impedance of air.

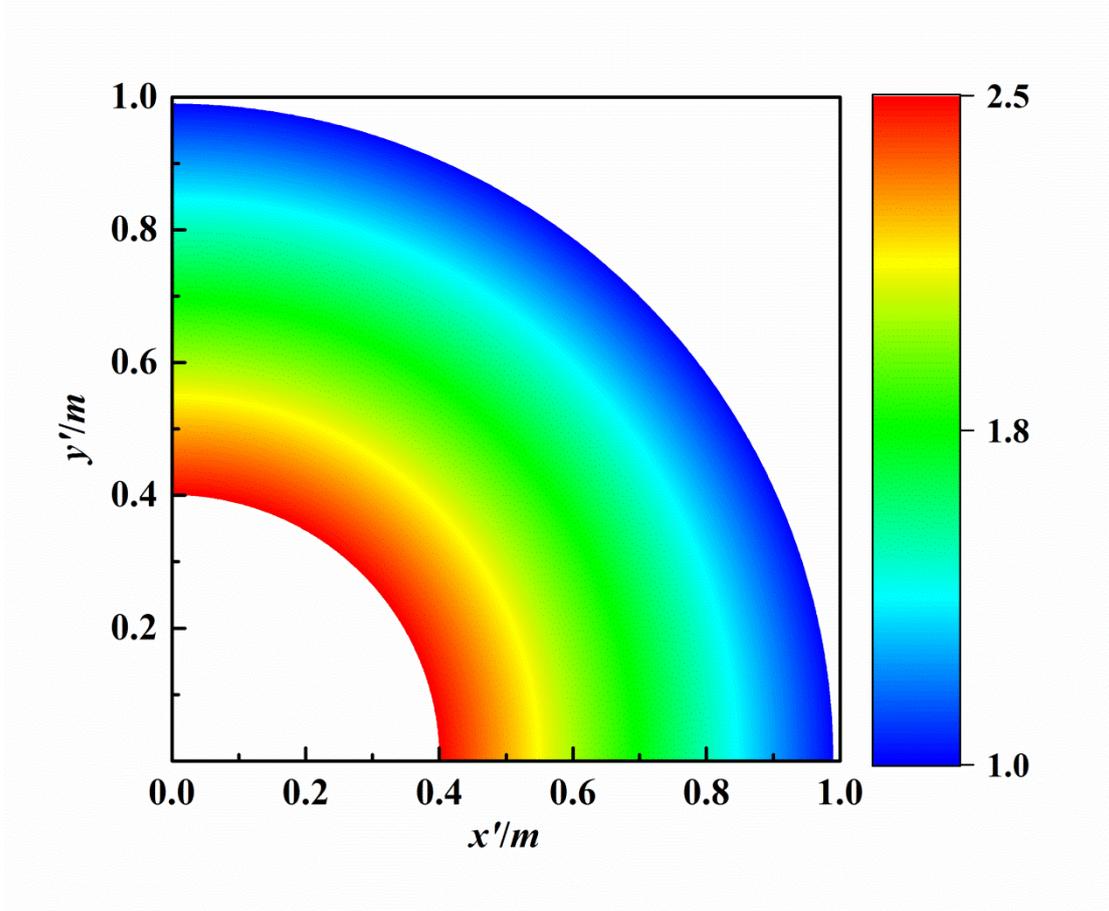

Figure 3. The spatial distribution of refractive index $n_\varphi'$ for the bend structure. The values of the refractive indices are represented by the colours.

From equation (5), one can observe that $n_r'$ is independent of the radius $r'$ while the refractive index $n_\varphi'$ is increasing with the decrease of the radius. The values range from 1 to 2.5. The distribution of refractive index $n_\varphi'$ for the bend structure is shown as a colour map in figure 3. The colours represent the values of the refractive indices.

Since the bending effect is decided mostly by the distribution of the refractive index, we focus on the profile of the refractive index. The impedance could be neglected without affecting the performance. To achieve the distribution of anisotropic refractive index for acoustic waves in equation (5), layers of perforated panel are used for the practical implementation.

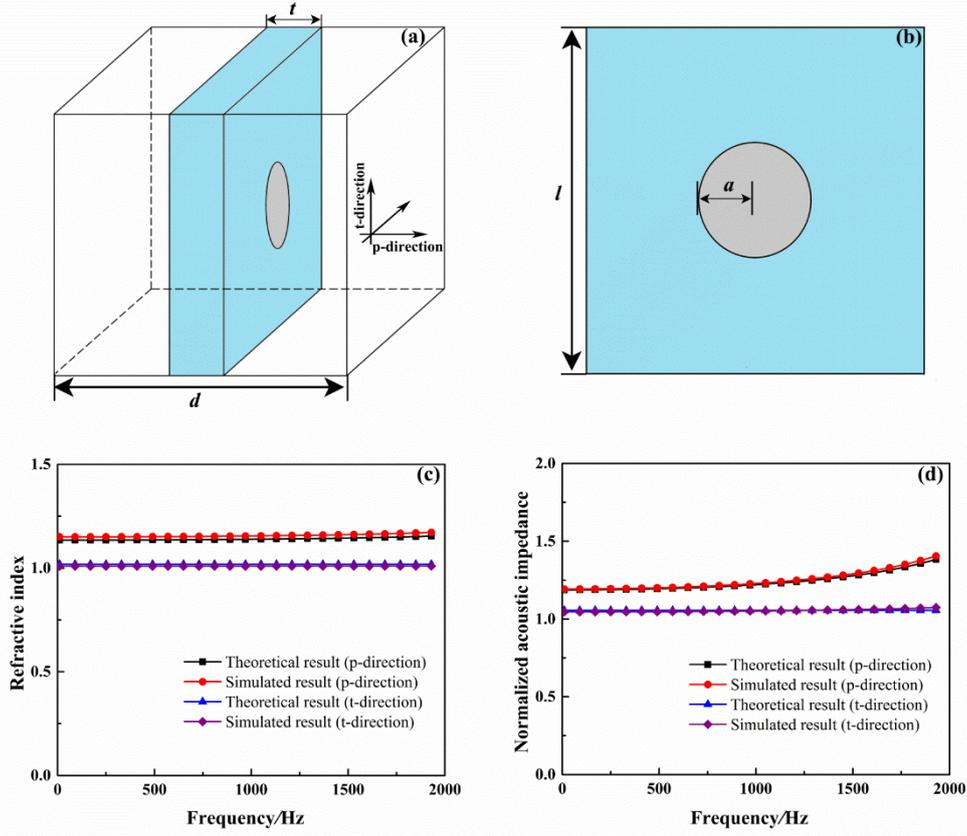

Figure 4. (a) Schematic representation of a typical unit cell. (b) Cross section of the typical unit cell. (c) Refractive indices and (d) the normalized acoustic impedances as functions of frequency for a typical unit cell. The geometric parameters of the typical unit cell is chosen as $t = 2mm$, $l = 10mm$, $a = 2.5mm$ and $d = 45mm$.

The model of a typical unit cell is given in figure 4(a) and its square cross section is illustrated in figure 4(b). The perforated panel with a hole is placed in the middle of the unit cell. The thickness $t$, the side length $l$ of the panel and the radius of the hole $a$ is 2 mm, 10 mm, 2.5 mm, respectively. The length of the unit cell $d$ is 45 mm, i.e., the thickness of the air layer is 43 mm. Besides, we define that the direction perpendicular to the plate is perpendicular direction (p-direction) and the direction parallel to the plate is tangential direction (t-direction), as shown in figure 4(a). Here both the theoretical model [21-25] and FEM [26] are used to obtain the effective acoustic parameters of the unit cell. The obtained effective refractive index and the normalized acoustic impedance are presented in figures 4(c) and (d). The results using these two methods fit well with each other. It can be observed that the anisotropic refractive index and normalized acoustic impedance are stable when the frequency is below 2 kHz.

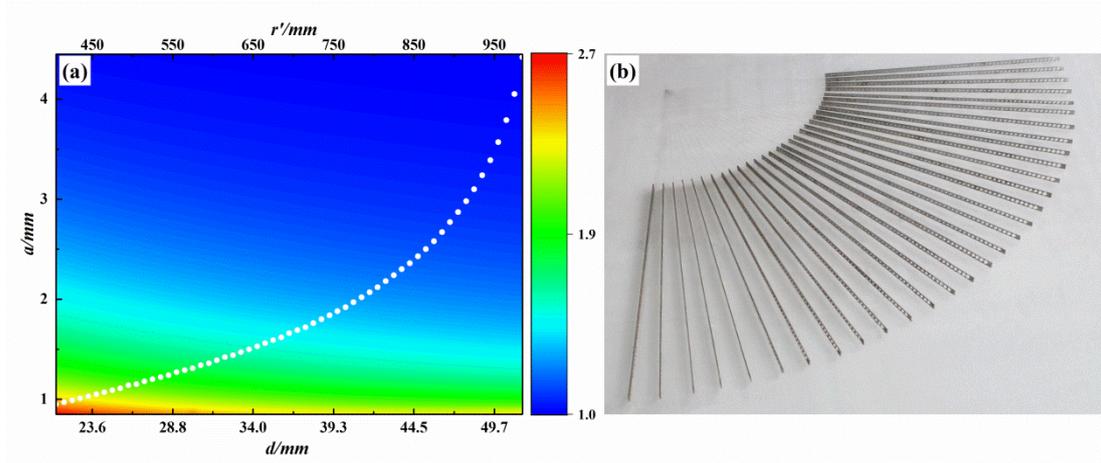

Figure 5. (a) The relationship for the refractive index with the length of the unit cell $d$ and the radius of the hole $a$ in the angular direction. The corresponding radical position $r'$ is also indicated as the top horizontal axis. (b) Photograph of the proposed structure sample.

It is known that the effective parameters of the unit cell change with the change of the geometric parameters. Therefore, we can obtain different effective refractive indices by adjusting the geometric parameters. The detailed relationship for the effective refractive index with the length of the unit cell $d$ (21.2-51.6 mm) and the hole-size $a$ (0.85-4.45 mm) of the perforated panel is calculated and presented as a colour map in figure 5(a). The obtained refractive index ranges from 1 to 2.7 shown in figure 5(a). So the required refractive index $n_\varphi'$, which ranges from 1 to 2.5, lies in a region which can be realized using the above unit cell.

Then, we constructed the model of acoustic right-angle bend shown in figure 5(b). It is composed of the same 31 strips along the angular direction. Each strip is made up of 59 unit cells with different radii of the holes. The strips are made of steel. Owing to the large impedance mismatch between the air and steel, the structured strips can be regarded as perfect rigid. Because the arc length increment between the adjacent unit cells is small, we could regard every single unit cell in the strip as a cuboid (figure 4(a)) whose length is the arc length of its centerline. Therefore, the length of the unit cell at the position of $r'$ can be expressed as follow: $d = \pi r'/60$. In addition, each unit cell can be considered as a homogeneous material with refractive index equal to that on its centerline. So figure 5(a) can also be regarded as the relationship for the refractive index with the position $r'$ (shown as the top horizontal axis in figure 5(a)) and the hole-size $a$ of the perforated panel. By adjusting the radius of the hole, different refractive indices can be obtained. Therefore, according to the required refractive index at different positions in equation (5), we can retrieve the corresponding radius of the hole shown as the white dots in figure 5(a). The hole-radius increases from 0.95 to 4.42mm in incremental steps from the inner to the outer along the $r'$-direction.

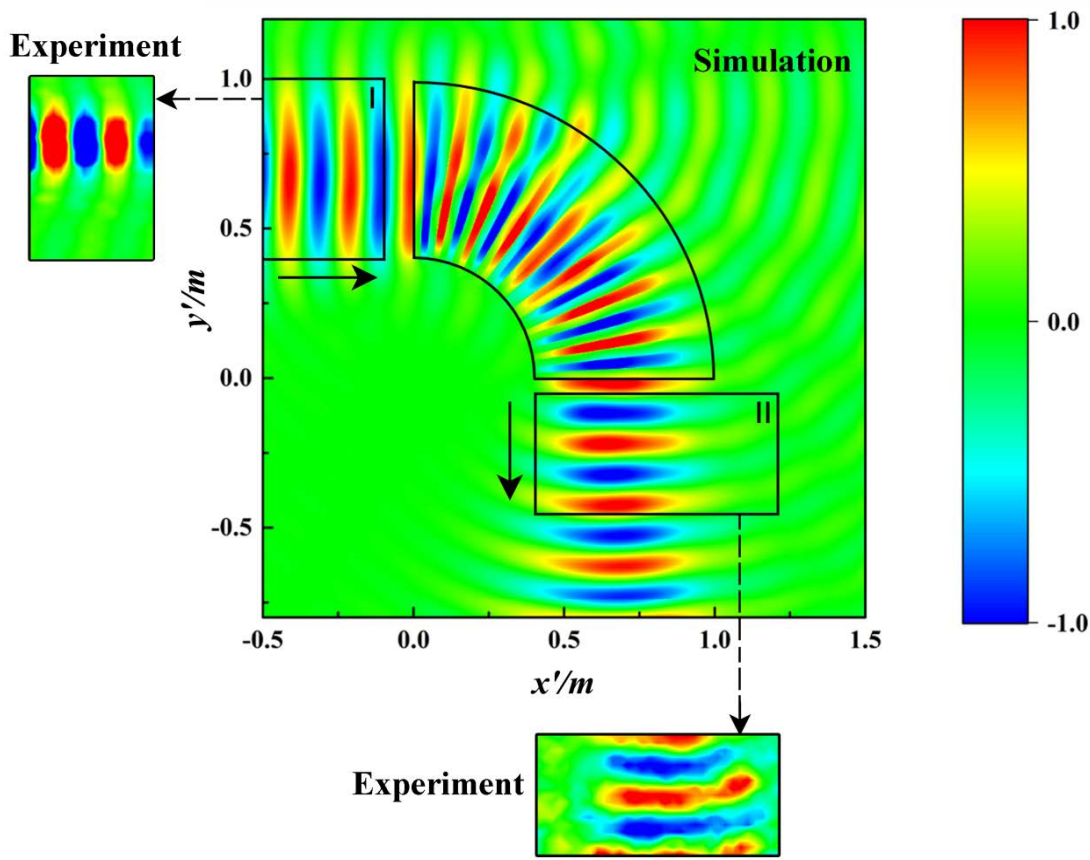

Figure 6. Simulated and measured sound pressure field distributions at 1700Hz. The field distributions are normalized to the maximum value.

Then, the effect of the bend made up of perforated panels is simulated at 1700 Hz. The structural parameters of the bend configured for the simulation were the same as those white dots in figure 5(a). The structure is placed in a 2D waveguide. The height of the waveguide is the same as the height of the perforated panel unit cell ($l$=10 mm). Therefore, the fields are confined in the channel region, where the wave fronts are planar surfaces and perpendicular to the propagation direction. The simulated field is presented in figure 6, which is normalized to the maximum value. The small black arrows represented the propagating direction. In simulation, the incident Gaussian beam propagated from left to right, and then entered the designed bend structure. Due to the distribution of the refractive index in the structure, the propagating direction started to change slowly. When the beam exited the bend structure, the propagating direction changed 90 degrees. The sound wave propagated from top to bottom while kept the Gaussian shape unchanged.

To demonstrate the performance of the device, experiments were also conducted. A 2D rectangular waveguide made up of two plexiglass plates was installed with an overall dimension of 2200 mm in length, 2000 mm in width, and 10 mm in thickness. The thickness of the waveguide is the same as the height of the perforated panel unit cell. The sound-absorbing cotton was put on the boundaries of the waveguide to minimize reflected waves. The fabricated acoustic right-angle bend was identical as configured for the simulation. In our experiment, an Audiobeam, which is an array made up of 260 mini transducers, was located on the left-hand side of the whole system as the sound source. It can generate an incident acoustic sound wave with strong directivity by utilizing the nonlinear propagation effects of ultrasonic in the air. A microphone (model: 46 B E, G. R. A. S.) is mounted on an auto scanner to measure the sound pressure field.

Two areas were measured to evaluate the performance of the bend, which were marked as area Ⅰ and area Ⅱ. The scanned incident area Ⅰ (entrance) is $400mm \times 600mm$ and the transmitting area Ⅱ (exit) is $800mm \times 400mm$ as shown in figure 6. The performance of our acoustic right-angle bend is examined from 1000 to 2000 Hz. The measured sound pressure field distributions for the two areas at 1700 Hz are presented in figure 6 to make a comparison with the simulated acoustic pressure field. From the measured acoustic pressure field in area Ⅰ, the incident wave was a high-directivity beam which propagated from left to right. While in area Ⅱ, despite of some diffusions, the propagating direction was from top to bottom, which was successfully rotated by 90 degrees. Obviously, the beam is smoothly bent along the circumferential direction both in simulation and experiment and these results agree well with theoretical predictions. Comparing simulations with the experimental results in area Ⅰ and Ⅱ, we can observe that the sound phases also coincide with each other.

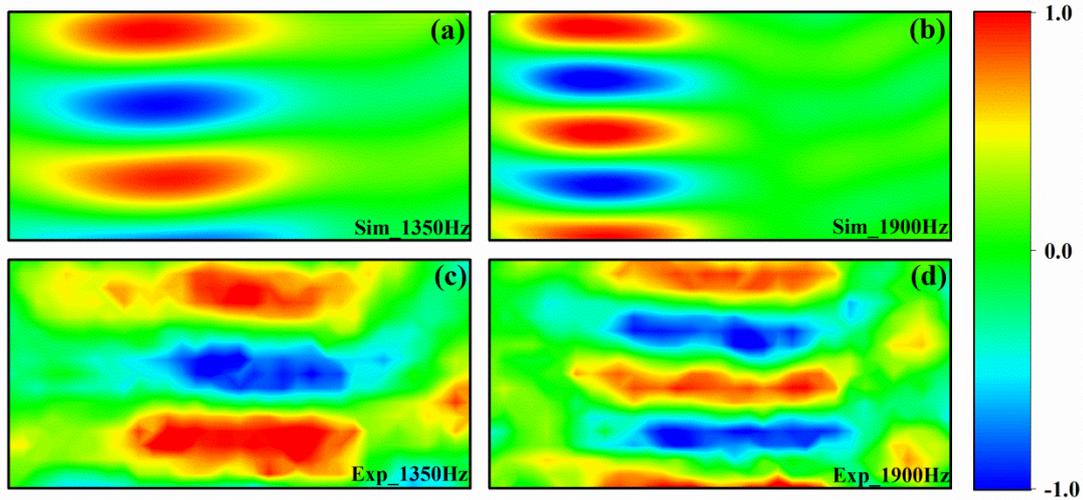

Figure 7. Simulated (upper panels) and measured (under panels) sound pressure distributions of the area Ⅱ for 1350Hz and 1900Hz. All data are normalized to the maximum value.

Furthermore, similar acoustic pressure fields in area Ⅱ from simulation and experiment are also provided when the operation frequencies are 1350 Hz and 1900 Hz, as shown in figure 7. The left two plots are the normalized acoustic pressure fields for 1350 Hz, while the right two are the normalized acoustic pressure fields for 1900 Hz. These results confirm the validity of the right-angle bend in a wide frequency range.

## 4. Conclusion

In conclusion, a 90-degrees acoustic bend structure consisting of perforated panels has been designed and experimentally demonstrated. The simulated and experimental results coincide with each other, which shows the effectiveness of the device. These results confirm that the acoustic wave bend can control the propagation of the acoustic wave in a broadband frequency range. This model may have potential for practical applications in various areas such as sound absorption and acoustic detection in pipeline

## Acknowledgments

This work is supported by the National Natural Science Foundation of China

(Grant No. 11304351) and the Youth Innovation Promotion Association CAS (Grant No. 2017029).# References

[1] Leonhardt U 2006 Optical conformal mapping *Science* **312** 1777

[2] Pendry J B, Schurig D and Smith D R 2006 Controlling electromagnetic fields *Science* **312** 1780

[3] Milton G W, Briane M and Willis J R 2006 On cloaking for elasticity and physical equations with a transformation invariant form *New Journal of Physics* **8** 248

[4] Greenleaf A, Lassas M and Uhlmann G 2003 Anisotropic conductivities that cannot be detected by EIT *Physiological measurement* **24** 413

[5] Chen H and Chan C T 2007 Acoustic cloaking in three dimensions using acoustic metamaterials *Applied physics letters* **91** 183518

[6] Cummer S A and Schurig D 2007 One path to acoustic cloaking *New Journal of Physics* **9** 45

[7] Norris A N. 2009 Acoustic metafluids *The Journal of the Acoustical Society of America* **125** 839

[8] Cummer S A, Rahm M and Schurig D 2008 Material parameters and vector scaling in transformation acoustics *New Journal of Physics* **10** 115025

[9] Li R Q, Zhu X F, Liang B, Li Y, Zou X Y and Cheng J C 2011 A broadband acoustic omnidirectional absorber comprising positive-index materials *Applied Physics Letters* **99** 193507

[10] Zigoneanu L, Popa B I and Cummer S A 2011 Design and measurements of a broadband two-dimensional acoustic lens *Physical Review B* **84** 024305

[11] Park J J, Lee K J B, Wright O B, Jung M K and Lee S H 2013 Giant Acoustic Concentration by Extraordinary Transmission in Zero-Mass Metamaterials *Physical Review Letters* **110** 244302

[12] Cheng Y, Zhou C, Wei Q, Wu D J and Liu X J 2013 Acoustic subwavelength imaging of subsurface objects with acoustic resonant metalens *Applied Physics Letters* **103** 224104

[13] Popa B I and Cummer S A 2009 Design and characterization of broadband acoustic composite metamaterials *Physical Review B* **80** 174303
13